\documentstyle[revtex]{aps}
\begin{document}
\draft
\begin{title}
High-Temperature series for the $RP^{n-1}$ lattice spin model\\
(generalized Maier-Saupe model of nematic liquid crystals)\\
in  two space dimensions and with general spin dimensionality $n$
\end{title}
\author{P. Butera and M. Comi}
\begin{instit}
Istituto Nazionale di Fisica Nucleare\\
Dipartimento di Fisica, Universit\`a di Milano\\
Via Celoria 16, 20133 Milano, Italy
\end{instit}
\begin{abstract}
High temperature series expansions of
the spin-spin correlation
functions of the $RP^{n-1}$  spin model on the square lattice
are computed through order
$\beta^{8}$  for general spin dimensionality
$n$.

Tables are reported for the expansion coefficients of
the energy per site, the susceptibility
and the second correlation moment. \end{abstract}
\widetext
\section{ Introduction }

Interest in the classical $RP^{n-1}$ spin systems
[\cite{maier}]
on a two-dimensional
lattice has been revived recently by
the results of a MonteCarlo
simulation[\cite{kunz}]  interpreted as evidence
of a second order "topological"
 phase transition, taking place for
values of the spin dimensionality
$n \geq 3$.
This is unexpected according to
renormalization group ideas.
Indeed the $RP^{n-1}$ models have the same
formal continuum limit   as
the conventional $O(n)$ symmetric $n-$vector  spin
models,  therefore they should
belong to the same universality
class and  should not behave
differently for $n \geq 3$
( when  $n=2$ the $RP^{n-1}$ model trivially
reduces to the  $n-$vector model).
However the global topologies of the
spin manifolds: the hypersphere $S^{n-1}$
 with antipodal points identified in
the case of the $RP^{n-1}$ model and simply
$S^{n-1}$ in the case of the $n$-vector
model, are different and it has long been known
that this might
be a reason for different phase diagrams[\cite{kogut}].

 MonteCarlo studies of these systems,
 mainly in the $n=3$ case,
sometimes with conflicting or not
completely convincing results are by now numerous
[\cite{mountain,duane,solomon,fukugita,sinclair,case,chiccoli}], and
they have been augmented
 by  recent more   extensive  simulations
on large lattices [\cite{kunz,wolff,caracciolo}]
using cluster algorithms[\cite{sokal}] in order to reduce the
critical slowing-down.
On the other hand, high temperature expansion (HTE) studies are still
practically absent, the only exceptions
being, to the best of our
knowledge, a series through order $\beta^{9}$ for
the internal energy
 and a series for the mass gap through order
$\beta^{5}$ in the $n=3$ case
[\cite{fukugita}].
These expansions have been helpful for
a first check of MonteCarlo
simulation codes, and series for
other quantities and for other values
of $n$ would be equally welcome.

We have extended to every value of the spin dimensionality $n$
through order $\beta^{8}$
 the computation of the internal energy ,
 and for the first time we have computed
series for the susceptibility and
the second correlation moment.

These series are probably not  long enough to provide, by
 their own, convincing evidence about the
existence, the location and
 the nature of a possible critical point, but
we believe it is useful to make them promptly
available so that they
can  serve  not only to check MonteCarlo data,
but also for future more extensive high
temperature calculations.

We shall explain later why our computational method,
 based on the Schwinger-Dyson
recursion equations[\cite{bcm}],
although very transparent, becomes rapidly
cumbersome and therefore is
unable, in its present form, to produce
substantially longer series.
\section{ The High Temperature Series }
Let us briefly describe the model and fix our conventions.

The partition function of the model is
\FL
\begin{eqnarray}
Z = \int \prod_{x}ds(x)\delta(s(x)^2-1)
 exp[\frac {\beta}{2}\sum_{x}\sum_{\mu=1,2}
(s(x)\cdot s(x+e_{\mu}))^{2}]
\end{eqnarray}
The variables of the model are
$n$-component classical spins $s(x)$ of unit
length associated to each site $x = x_{1}e_{1} + x_{2}e_{2}$
of a 2-dimensional square lattice,
 $e_{1}$ and  $e_{2}$ are the two elementary
lattice vectors.

The Hamiltonian and the integration measure have
a global $O(n)/Z_{2}$ and a local
$Z_{2}$ invariance.
Since in two dimensions continuous symmetries are
unbroken[\cite{mermin}], the most
general correlation function $<\phi(C)>$ can be written as
\begin{eqnarray}
<\phi(C)> = <\phi( x_1,x_2,...,x_n; \{b_{i,j}\})>
=<\prod_{1 \leq i <j \leq n}
(s(x_{i}) \cdot s(x_{j}))^{b_{ij}}>
\label{eq:correl} \end{eqnarray}
with integer $b_{i,j} \geq 0 $.
The local invariance under  $Z_{2}$, which also
cannot break[\cite{elitzur}],
implies the further restriction that
each $s(x_i)$ has to appear in $\phi(C)$
an even number of times.

The correlation function (\ref{eq:correl})  may
be represented graphically
 as follows:
the lattice points $x_1,x_2,...,x_n$ are taken
as vertices and  a line connecting
the vertices $x_i$ and $x_j$ is associated to
each factor $s(x_i) \cdot s(x_j)$ in $\phi(C)$.
In terms of graphs the local $Z_{2}$ invariance requires that
the degree of each vertex be even.
Thus, for instance, the correlation
$<s(x_1) \cdot s(x_2)>$ vanishes trivially.

The fundamental two-spin correlation is then
$ G(x_2-x_1;\beta, n)=<(s(x_1) \cdot s(x_2))^{2}>$.

In particular we have $-  G(e_{1};\beta, n) = E$,
the energy per site.

We also have computed the moments
$m^{(l)}(\beta, n)$ of the connected
correlations
\FL
\begin{equation}
C(x_2-x_1;\beta, n) =
\sum_{a,b}<\Big (s^{a}(x_1) s^{b}(x_1)
-<s^{a}(x_1) s^{b}(x_1)>\Big)
\Big(s^{a}(x_2) s^{b}(x_2)-<s^{a}(x_2) s^{b}(x_2)>\Big)>
= G(x_2-x_1;\beta, n)- 1/n
\end{equation}

which are defined as follows
\begin{equation}
m^{(l)}(\beta, n) = \sum_{x}\mid x \mid^{l} C(x;\beta,n)
=\sum_{r}a_{r}^{(l)}\beta^{r}
\end{equation}

The HTE coefficients for  $G(e_{1};\beta, n)
=\sum_{r}g_{r}(n)\beta^{r}$ are:
\begin{eqnarray*}
g_{0}(n) =\frac{1}{n}
\end{eqnarray*}
\begin{eqnarray*}
g_{1}(n) =\frac{n-1}{n^2 (n+2)}
\end{eqnarray*}
\begin{eqnarray*}
g_{2}(n) =\frac{(n-1)(n-2)}{n^3 (n+2)(n+4)}
\end{eqnarray*}
\begin{eqnarray*}
g_{3}(n) =\frac{(n-1)(72+18n-11n^2-n^3+n^4)}
{n^4 (n+2)^3 (n+4) (n+6)}
\end{eqnarray*}
\widetext
\begin{eqnarray*}
g_{4}(n) =\frac{(n-1)(n-2)(528+130n-17n^2-3n^3+n^4)}
{n^5 (n+2)^3 (n+4)(n+6)(n+8)}
\end{eqnarray*}
\begin{eqnarray*}
g_{5}(n) =((n-1)(284160+130496n-104032n^2-53344n^3+6888n^4
+5496n^5+474n^6-56n^7-2n^8+n^9)\\
 /(n^6 (n+2)^5 (n+4)^2 (n+6) (n+8) (n+10))
\end{eqnarray*}
\begin{eqnarray*}
g_{6}(n) =(n-1)(n-2)
   (11704320+8093952n-1233088n^2-1863104n^3
-200776n^4+103840n^5\\
+26210n^6+1386n^7-100n^8-2n^9+n^{10})
 /(n^7 (n+2)^5 (n+4)^3 (n+6) (n+8) (n+10) (n+12))
\end{eqnarray*}
\FL
\begin{eqnarray*}
g_{7}(n) =(n-1)
  (341118812160+428301582336n+17644511232n^2
-191549657088n^3
   -76694446080n^4\\
+17276826240n^5
+16424658272n^6+1926697808n^7
   -951227456n^8-295105184n^9\\
-5505626n^{10}
+10001781n^{11}
   +1876337n^{12}+133277n^{13}+1527n^{14}
-171n^{15}+9n^{16}+n^{17})\\
 /(n^8(n+2)^7(n+4)^4(n+6)^3(n+8)(n+10)(n+12)(n+14))\\
\end{eqnarray*}
\FL
\begin{eqnarray*}
g_{8}(n) =(n-1)  (n-2)
   (1271577968640 + 1237547925504n - 87404783616n^2
- 441393059328n^3\\
 - 107082739328n^4
+ 37546256480n^5 +  17834481104n^6
+ 440575008n^7 - 777645296n^8\\
-   105547274n^9 + 10134853n^{10}
+ 3591697n^{11}
 + 316891n^{12}  + 7749n^{13}
- 289n^{14} - n^{15} + n^{16} )\\
/(n^9(n+2)^7(n+4)^3(n+6)^3(n+8)(n+10)(n+12)(n+14)(n+16))\\
\end{eqnarray*}
For $n=3$ we have (compare with Ref.[\cite{fukugita}]):
\begin{eqnarray*}
G(e_1;\beta, 3)=\frac {1} {3}+\frac{2} {45}\beta
+\frac{2} {945}\beta^2+\frac{2} {7875}\beta^3
+\frac{34} {467775}\beta^4
+\frac{13402} {2280403125}\beta^5
+\frac{10702} {47888465625}\beta^6\\
+\frac{12179386} {142468185234375}\beta^7
+\frac{33996598} {4872411935015625}\beta^8+...\\
\end{eqnarray*}

The HTE coefficients for
$m^{(0)}(\beta, n)$, also called the
susceptibility, are:
\begin{eqnarray*}
a_{0}^{(0)}(n)=\frac{n-1}{n}\\
\end{eqnarray*}
\begin{eqnarray*}
a_{1}^{(0)}(n)=\frac{4(n-1)}{n^2(2 + n)}\\
\end{eqnarray*}
\begin{eqnarray*}
a_{2}^{(0)}(n)=\frac{4(n-1)(8 + 3n + n^2)}
{n^3(2 + n)^2(4 + n)}\\
\end{eqnarray*}
\begin{eqnarray*}
a_{3}^{(0)}(n)=\frac {4(n-1)(96 + 64n + 32n^2 + 5n^3 + n^4)}
{n^4(2 + n)^3(4 + n)(6 + n)}\\
\end{eqnarray*}
\widetext
\begin{eqnarray*}
a_{4}^{(0)}(n)=
\frac{4(n-1)(1 + n)(3456 + 1968n + 570n^2
+ 89n^3 + 9n^4 + n^5)}
  {n^4(2 + n)^4(4 + n)^2(6 + n)(8 + n)}\\
\end{eqnarray*}
\begin{eqnarray*}
a_{5}^{(0)}(n)=
(4(n-1)(122880 + 101888n + 40640n^2 + 42528n^3 + 35696n^4 + 11094n^5\\
 + 1807n^6 + 162n^7 + 10n^8 + n^9))
  /(n^6(2 + n)^5(4 + n)^2(6 + n)(8 + n)(10 + n))\\
\end{eqnarray*}
\begin{eqnarray*}
a_{6}^{(0)}(n)=
(4(n-1)(-115015680 - 79331328n + 74609664n^2 + 96772864n^3
 +  44006080n^4\\
+ 15702208n^5 + 7513312n^6
+ 2862016n^7 + 648560n^8
+  87178n^9 + 7048n^{10} + 364n^{11}
+ 19n^{12} + n^{13}))\\
/(n^7(2 + n)^6(4 + n)^3(6 + n)^2(8 + n)(10 + n)(12 + n))\\
\end{eqnarray*}
\FL
\begin{eqnarray*}
a_{7}^{(0)}(n)=
(4(n-1)(43104337920 + 43866980352n
- 5407064064n^2 - 15002345472n^3 +  3765867520n^4\\
+ 8878097920n^5
+ 4282305280n^6 + 1196842912n^7
+  326380672n^8 + 97376320n^9\\
+ 22123168n^{10} + 3228422n^{11}
+  292472n^{12} + 16058n^{13}
+ 566n^{14} + 23n^{15} + n^{16}))\\
/(n^8(2 + n)^7(4 + n)^3(6 + n)^3(8 + n)
(10 + n)(12 + n)(14 + n))\\
\end{eqnarray*}
\begin{eqnarray*}
a_{8}^{(0)}(n)=
(4(n-1)(-94746307461120 - 126660674322432n
- 5226623926272n^2+  66792515567616n^3\\
+ 32795171340288n^4 + 48194863104n^5
 - 862398014464n^6 + 2921457334912n^7\\
+ 2239266005664n^8+  790758440112n^9 + 185029551696n^{10}
+ 37818452512n^{11} + 7776875970n^{12}\\
 + 1395247971n^{13} + 184588028n^{14} + 16678488n^{15}
+ 985722n^{16} + 36650n^{17} + 952n^{18}
+ 32n^{19} + n^{20}))\\
  /(n^9(2 + n)^8(4 + n)^4(6 + n)^3(8 + n)^2
(10 + n)(12 + n)(14 + n)(16 + n))
\end{eqnarray*}
\FL
For $n=3$ these formulae give:
\begin{eqnarray*}
m^{(0)}(\beta, 3)=\frac{2} {3}
+\frac{8} {45}\beta+\frac{208}
{4725}\beta^2
+\frac{704}{70875}\beta^3
+\frac{12704}{5457375}\beta^4
+\frac{8254816} {15962821875}\beta^5\\
+\frac{37545856}{335219259375}\beta^6
+\frac{10273872032}{427404555703125}\beta^7
+\frac{934133719808}{183909666174609375}\beta^8+...\\
\end{eqnarray*}

The HTE coefficients for
$m^{(2)}(\beta, n)$, the second  correlation moment, are:
\begin{eqnarray*}
a_{1}^{(2)}(n)=\frac{4(n-1)}{n^2(2 + n)}\\
\end{eqnarray*}
\begin{eqnarray*}
a_{2}^{(2)}(n)=\frac{4(n-1)(28 + 8n + n^2)}
{n^3(2 + n)^2(4 + n)}\\
\end{eqnarray*}
\begin{eqnarray*}
a_{3}^{(2)}(n)=\frac{4(n-1)(624 + 344n + 124n^2 + 15n^3
+ n^4)} {n^4(2 + n)^3(4 + n)(6 + n)}\\
\end{eqnarray*}
\widetext
\FL
\begin{eqnarray*}
a_{4}^{(2)}(n)=\frac{4(n-1)(52224 + 57856n
+ 37760n^2 + 13200n^3 + 2844n^4 + 376n^5 +
 25n^6 + n^7)}{n^5(2 + n)^4(4 + n)^2(6 + n)(8 + n)}\\
\end{eqnarray*}
\begin{eqnarray*}
a_{5}^{(2)}(n)= ( 4(n-1)
(1044480 + 1553408n + 1394176n^2 + 692768n^3
+ 237584n^4 + 50998n^5\\
  + 7107n^6 + 642n^7 + 30n^8 + n^9))
  / ( n^6(2 + n)^5(4 + n)^2(6 + n)(8 + n)(10 + n))\\
\end{eqnarray*}
\begin{eqnarray*}
a_{6}^{(2)}(n)=
(4(n-1)(420249600 + 1092464640n
+ 1523555328n^2 + 1255855616n^3+ 710497728n^4\\
+ 285572064n^5 + 84264528n^6 + 18404144n^7 + 2909092n^8
 + 327054n^9 + 25803n^{10} + 1354n^{11} + 44n^{12} + n^{13}))\\
/( n^7(2 + n)^6(4 + n)^3(6 + n)^2
(8 + n)(10 + n)(12 + n))\\
\end{eqnarray*}
\begin{eqnarray*}
a_{7}^{(2)}(n)=(4(n-1)(83979141120
+ 204309430272n + 291728203776n^2 +  300109848576n^3\\
+ 232510854656n^4 + 131053062400n^5
 + 54800469376n^6 + 17324337248n^7 + 4212618016n^8\\
+ 794772080n^9  + 114865432n^{10} + 12396858n^{11} + 977532n^{12}
+  55028n^{13} + 2106n^{14} + 53n^{15} + n^{16}))\\
/(n^8(2 + n)^7(4 + n)^3(6 + n)^3(8 + n)
(10 + n)(12 + n)(14 + n))\\
\end{eqnarray*}
\begin{eqnarray*}
a_{8}^{(2)}(n)=(4(n-1)
 (-58788371496960 + 8364704661504n + 280171118592000n^2
+ 479504520511488n^3\\
 + 464202911416320n^4 + 329227765829632n^5
+ 188047485044736n^6 + 87029638424064n^7\\
+ 32173730443520n^8
+ 9456558685824n^9 + 2219800018368n^{10}
+ 419325652576n^{11}  + 63930454192n^{12}\\
+ 7800108776n^{13}
+ 746995212n^{14} + 54841620n^{15}
+ 3013992n^{16} + 120454n^{17} + 3374n^{18}
+ 67n^{19} + n^{20}))\\
/(n^9(2 + n)^8(4 + n)^4(6 + n)^3
(8 + n)^2(10 + n)(12 + n)(14 + n)
(16 + n))\\
\end{eqnarray*}
In particular, the HT expansion of $m^{(2)}(\beta, 3)$ is:
\FL
\begin{eqnarray*}
m^{(2)}(\beta, 3)=\frac{8}{45}\beta+\frac{488} {4725}\beta^2
+\frac{2896}{70875}\beta^3
+\frac{1123712}{81860625}\beta^4
+\frac{67018144}{15962821875}\beta^5
+\frac{2023066384}{1676096296875}\beta^6\\
+\frac{12824336768}{38854959609375}\beta^7
+\frac{18110407484144}{208430954997890625}\beta^8+...
\end{eqnarray*}

A correlation length may be defined, as
usual, in terms of the ratio
 of $m^{(2)}(\beta, n)$ and $m^{(0)}(\beta, n)$.

Let us notice that a few simple checks
of the formulae are possible:
 all HTE of the connected correlations
have to vanish for $n=1$ because of
the triviality of the $RP^{0}$
 model. For $n=2$, the expansions
should reduce to the corresponding
ones for the $O(2)$ (or XY-) vector model. Finally,
for $n=3$ the HTE of $C(e_{1};\beta, n)$
agrees with the calculation of Ref.[\cite{fukugita}].

Our HTE have been computed from
the Schwinger-Dyson equations of the model,
an infinite system of linear equations among the
correlation functions.
The generic equation, which may be deduced following
closely Ref.[\cite{bcm}], has the structure
\FL
\begin{eqnarray}
<\phi(C)> = \frac {1} {n + g_1-2}
\Big [ \beta\sum_{\mu}(<\phi(C^{-}_{\mu})>-
<\phi(C^{+}_{\mu})>)\nonumber
+(b_{12}-1)<\phi(C_{12,12})>-
 \sum_{j=3}^n b_{1j}
<\phi(C^{2j}_{12,1j})> \Big ]\nonumber\\
\label{eq:sde} \end{eqnarray}
Here  we have assumed that the vertices
 $x_{1}$ and $x_{2}$ are connected
 by one line at least,
$g_{1}$ is the degree of the
vertex $x_{1}$, $b_{ij}$ the number of
lines connecting the vertices  $x_{i}$ and $x_{j}$,
$<\phi(C^{-}_{\mu})>$ denotes the correlation function
 obtained from $<\phi(C)>$ by removing a
factor $s(x_1) \cdot s(x_2)$
 and replacing it by
$s(x_1) \cdot s(x_{1+\mu})s(x_2) \cdot s(x_{1+\mu})$,
namely
\begin{equation}
\phi(C^{-}_{\mu})= \phi(C)
\frac{s(x_1) \cdot s(x_{1+\mu})s(x_2) \cdot s(x_{1+\mu})}
{s(x_1) \cdot s(x_2)}\nonumber\\
\end{equation}
and analogously
\begin{eqnarray*}
\phi(C^{+}_{\mu})&&=\phi(C)
( (s(x_1) \cdot s(x_{1+\mu}))^{2}\nonumber\\
\phi(C_{12,12})&&=\frac{\phi(C)}
{(s(x_1) \cdot s(x_2))^{2}}\nonumber\\
\phi(C^{2j}_{12,1j})&&= \phi(C)
\frac{s(x_2) \cdot s(x_{j})}{s(x_1)
\cdot s(x_2)s(x_1) \cdot s(x_j)}.\nonumber\\
\end{eqnarray*}
The HTE of the correlation $<\phi(C)>$
is obtained  solving iteratively
eqs.(\ref{eq:sde}) by the same procedure as
in the case of the $n-$vector model[\cite{bcm}].
Here  however, a difficulty is met:
while in the case of the $n-$vector model a
large fraction of the
graphs generated after the first
few iterations can be neglected,
in this case, due to the local $Z_2$ symmetry, all
graphs contribute nontrivially
to the final results and
therefore must be recorded. Thus the required
computer memory rapidly becomes  exceedingly
large and it is difficult to
push the expansion beyond the 8-th order.
However not all the blame should be laid upon
the computational technique since the combinatorial
complexity of the expansion is really
higher and of a faster
growth with the order than in the $n-$vector
case. It is also interesting
to recall that analogous difficulties
were met when performing  strong coupling
expansions in the Hamiltonian
formalism[\cite{solomon}].

A simple analysis of the series by
ratio  and Pad\'e
approximants methods[\cite{seran}] (see Fig.1 and Fig.2 )
 suggests the existence of a
critical point when $ n \approx 2$, but,
unfortunately, the series
seem to be not long enough to warrant
any reasonably safe conclusion when
$n = 3$ or greater.
To be  sure, for various values of
$n$ there are some Pad\'e
approximants of the susceptibility
 having a real positive singularity or a
complex conjugate pair
of singularities nearby the real positive $\beta$ axis and
in the expected position. The same
happens for the
logarithmic derivative of the susceptibility.
These poles  however, at this order of approximation, are not
stable enough  to
enable us to exclude the possibility of
an artifact of low
order approximants to mimic the steep
increase of the susceptibility.
Thus some completely different scenarios are
still compatible
with our series, for instance:

a) in analogy with the behavior of the $n-$vector
model[\cite{bcm}] a  critical point exist for $n \approx 2$.
As $n$ is increased and varied through some
$\tilde n \leq 3$, the critical point might
split into an
unphysical pair of complex conjugate
singularities
so that the model becomes asymptotically free
for $n \geq 3$.
This conjecture might be
supported both by the alternate ratios plots
of Fig. 1, which seem to show the
onset of an oscillatory trend [\cite{seran}]
and by some Pad\'e approximants
to the susceptibility or its log-derivative
whose nearest singularities
in the right half $\beta$ plane are  complex.

b)  a  critical point exists for all $n$
as suggested by Ref.[\cite{kunz}].
\nonum
\section{Acknowledgments}
Our thanks are due to Sergio Caracciolo for
suggesting to undertake this
computation and to Alan Sokal
for further encouragement and useful
discussions. We also are indebted to U. Wolff
for a useful discussion and
for kindly permitting us to use his
unpublished MonteCarlo data in Fig.2.
Finally we thank A. J. Guttmann and
G. Marchesini for carefully reading a draft of this note.
Our work has been partially supported
by MURST.
\narrowtext

\figure{Alternate ratios $\bar r_{s}(n)
= (a_{s-1}^{(0)}(n)/a_{s+1}^{(0)}(n))^{1/
2}$  of the expansion coefficients of the susceptibility
for various values of the spin dimensionality $n$
are plotted versus $1/s$.
Going from the lower plot to the upper,
we have $n=2,3,..,8$.\label{ratioplot}}
\figure{ The susceptibility for $n=3$
at order $\beta^{8}$ vs. $\beta$.
The continuous curve shows
the $[4/4]$ Pad\'e approximant
(which is singular at
$\beta \approx 5.175 \pm0.315i$).
The dashed curve shows the sum of
the susceptibility series truncated at
order $\beta^{8}$.
The squares represent data from the
MonteCarlo simulation of Ref.
[\cite{fukugita}]. The triangles represent
unpublished
data from a MonteCarlo cluster simulation
performed by U. Wolff[\cite{wolff}]
on lattices of size up to $256^2$.\label{sus}}
\end{document}